\documentstyle[aps,prl,twocolumn]{revtex}

\def\be{\begin{equation}}
\def\ee{\end{equation}}
\def\bea{\begin{eqnarray}}
\def\eea{\end{eqnarray}}
\def\bma{\begin{mathletters}}
\def\ema{\end{mathletters}}

\def\0{\overline{0}}

\def\q0{\underline{0}}

\def\C{{\cal C}}
\def\E{{\cal E}}

\def\tr{\mbox{tr}}

\def\bra#1{\langle#1|} \def\ket#1{|#1\rangle}
\def\braket#1#2{\langle#1|#2\rangle}

\def\proj#1{\ket{#1}\!\bra{#1}}

\begin{document}

\draft \wideabs{

\title{
On the structure of a reversible entanglement generating set for
three--partite states }

\author{A. Ac\'\i n$^{1*}$,
G. Vidal$^{2\dagger}$ and J. I. Cirac$^{3\ddagger}$}

\address{
$^{1}$GAP-Optique, University of Geneva, 20, Rue de l'\'Ecole de M\'edecine, CH-1211 Geneva 4, Switzerland\\
$^{2}$Institute for Quantum Information, California Institute of Technology, Pasadena, CA 91125, USA\\
$^{3}$Max--Planck Institut f\"ur Quantenoptik, Hans--Kopfermann
Str. 1, D-85748 Garching, Germany }

\date{\today}

\maketitle

%%%%%%%%%%%% Abstract %%%%%%%%%%%%%%%%%%%%%%%%%%%

\begin{abstract}

We show that Einstein--Podolsky--Rosen--Bohm (EPR) and
Greenberger--Horne--Zeilinger--Mermin (GHZ) states can not
generate, through local manipulation and in the asymptotic limit,
all forms of three--partite pure--state entanglement in a
reversible way. The techniques that we use suggest that there may
be a connection between this result and the irreversibility that
occurs in the asymptotic preparation and distillation of bipartite
mixed states.

\end{abstract}

}

To identify the fundamentally inequivalent ways quantum systems
can be entangled is a major goal of quantum information theory. In
the case of systems shared by two parties, Alice and Bob, there is
only one type of entanglement, namely that contained in the
Einstein--Podolsky--Rosen--Bohm (EPR) state \be \label{singlet}
\ket{EPR}=\frac{1}{\sqrt 2}\left(\ket{00}+\ket{11}\right) , \ee in
the sense that, in the limit of large $N$, Alice and Bob can {\em
reversibly} transform $N$ copies of any other state
$\ket{\psi}_{AB}$ into EPR states by using only local operations
and classical communication (LOCC) \cite{BBPS}. This simple
picture becomes much richer in systems shared by more than two
parties, since also genuine multipartite entanglement exists
\cite{BPRST}. In particular, the
Greenberger--Horne--Zeilinger--Mermin (GHZ) state \be
\label{GHZst} \ket{GHZ}=\frac{1}{\sqrt
2}\left(\ket{000}+\ket{111}\right) \ee can not be reversibly
generated from EPR states pairwise distributed among Alice, Bob
and a third partie Claire \cite{LPSW}. In the terminology of Ref.
\cite{BPRST}, this implies that EPR states alone do not form a
minimal reversible entanglement generating set (MREGS) for
three--partite states.

The results of Ref. \cite{LPSW} left open the question whether,
instead, the set \be \label{3mregs}
G_3=\{\ket{GHZ},\ket{EPR}_{AB},\ket{EPR}_{AC},\ket{EPR}_{BC}\} \ee
constitutes a MREGS. Denoting by $\rightleftharpoons$ an
asymptotically (i.e. in the large $N$ limit) reversible
transformation using LOCC, this question amounts to assessing the
feasibility of a transformation of the form \bea
\ket{\psi}_{ABC}^{\otimes N} \ \ \rightleftharpoons &&
\ket{GHZ}^{\otimes gN} \otimes \ket{EPR}_{AB}^{\otimes xN} \label{trans}\\
&&\otimes \ket{EPR}_{AC}^{\otimes yN} \otimes
\ket{EPR}_{BC}^{\otimes zN}, \nonumber \eea where $g,x,y,z \geq
0$, for any three--partite state $\ket{\psi}_{ABC}$. If this were
the case, then entanglement in three--partite systems could be
regarded as consisting only of GHZ and EPR correlations.

In the meantime it has been proved that not all four--partite
states can be reversibly generated from a distribution of EPR and
three-- and four--partite GHZ states \cite{WZ}. However, no
evidence has been found contradicting the following conjecture.

\vspace{1mm}

{\bf Conjecture:} $G_3$ is a MREGS for three--partite states.

\vspace{1mm}

On the contrary, all reversible transformations of three--partite
states so far reported, involving Schmidt decomposable states
\cite{BPRST}, but also a whole class of more elaborated states
\cite{VDC}, seem to support it.

In this Letter we give examples of three--partite states, denoted
by $\ket{\Psi_\delta}_{ABC}$, that can not be reversibly generated
only with states of the set $G_3$, thus disproving the above
conjecture. We also show that even a reversible transformation of
states of $G_3$ into any of these states {\em and} states of $G_3$
is impossible. That is, we show that there are cases where the
transformation of Eq. (\ref{trans}) can not be made reversible
even if the coefficients $g,x,y,z$ are eventually allowed to be
negative \cite{explain1}. Notice that such a possibility, not
previously excluded in four--partite systems, would have allowed
for a slightly different description of multipartite entanglement,
also based exclusively on EPR and GHZ correlations.

These results, therefore, indicate the need to extend the set
$G_3$ in order to eventually obtain a MREGS, either in its
original formulation or in the extended sense described above. We
would like to note, however, that the notion of a non--trivial
MREGS implicitly assumes that the manipulation of multipartite
pure states can be made reversible. This is, admittedly, an
appealing idea, but has not yet been proved. In this sense, our
results can be just interpreted as to indicate that a fundamental
irreversibility occurs during the process of combining EPR and GHZ
entanglements into any of the three--partite pure states
$\ket{\Psi_\delta}$.

It is natural to inquire into the origin of such an
irreversibility, which is somewhat analogous to the one that
characterizes the cycle of preparing and distilling bipartite
mixed states \cite{VC}. Actually, the argument that will lead to
disprove the above conjecture would fail if mixed--state
entanglement could be reversibly distilled. This fact suggests a
connection between the two irreversible processes.

Our strategy consists in showing that a conservation law obeyed in
reversible asymptotic entanglement transformations \cite{LPSW}
would be violated if EPR and GHZ states could generate
$\ket{\Psi_\delta}$ reversibly. Let $\ket{\Psi}_{ABC}$ denote an
arbitrary three--partite pure state shared by Alice, Bob and
Claire, and let $\rho_{AB}$ be the mixed state resulting from
tracing out Claire's subsystem. The relative entropy of
entanglement of $\rho_{AB}$ \cite{relent}, \be \label{ree}
E_{\Omega}(\rho_{AB})\equiv\min_{\sigma_{AB}\in \Omega}
S(\rho_{AB}\,||\,\sigma_{AB}), \ee where $\Omega$ is some convex
set of states (typically, that of separable states) invariant
under LOCC and $S(\rho\,||\,\sigma)\equiv$
tr$(\rho\log_2\rho-\rho\log_2\sigma)$ is the quantum relative
entropy, was originally introduced to quantify the entanglement of
bipartite mixed states. Its regularized version, \be
E^{reg}_{\Omega}(\rho_{AB}) \equiv \lim_{ N \rightarrow \infty}
\frac{E_{\Omega}(\rho_{AB}^{\otimes N})}{N}, \ee is a lower bound
for the entanglement cost $E_c$ \cite{BDSW,cost} of $\rho_{AB}$,
or number of EPR states per copy of $\rho_{AB}$ needed to
asymptotically prepare copies of $\rho_{AB}$. It is also an upper
bound for its distillable entanglement $E_d$ \cite{BDSW,dist}, or
number of EPR states per copy of $\rho_{AB}$ that can be
asymptotically distilled from copies of $\rho_{AB}$. Indeed,
$E^{reg}_{\Omega}$ fulfills the postulates required in
\cite{limits} for an entanglement measure and therefore
\cite{limits,PV} \be E_c(\rho_{AB}) \ge
E^{reg}_{\Omega}(\rho_{AB}) \geq E_d(\rho_{AB}).
\label{inequalities} \ee Particularly relevant in the context of
this work will be the fact that, as showed in \cite{LPSW}, the
relative entropy of entanglement of (say) subsystems $AB$,
$E_\Omega (AB)$  must be conserved during any reversible
pure-state transformation of the system $ABC$. Applied to
transformation (\ref{trans}) this law reads \be E_{\Omega}
(\rho_{AB}^{\otimes N}) = E_{\Omega} ( [EPR]_{AB}^{\otimes xN}),
\ee $[EPR]\equiv \proj{EPR}$, where we have used that when tracing
out part $C$, only $\ket{EPR}_{AB}$ gives a non-separable
contribution \cite{explain2}. Thus, in the  large $N$ limit we are
left with the condition \be E^{reg}_{\Omega} (\rho_{AB}) = x,
\label{cond1} \ee where $x$ is the number of EPR states per copy
of $\rho_{AB}$ that should be available on the rhs of Eq.
(\ref{trans}), and we have used that $E_{\Omega}([EPR]_{AB}) = 1$.
Similarly, if instead we allow now for states of $G_3$ to appear
simultaneously on both sides of transformation (\ref{trans}), we
obtain \be E_{\Omega} (\rho_{AB}^{\otimes N}\otimes
[EPR]_{AB}^{\otimes x_1 N}) = E_{\Omega} ([EPR]_{AB}^{\otimes
x_2N}), \ee $x_1,x_2\geq 0$, which implies the condition \be
\lim_{N \rightarrow \infty} \frac{E_{\Omega}(\rho_{AB}^{\otimes
N}\otimes[EPR]_{AB}^{\otimes x_1 N})}{N} = x_2. \label{cond2} \ee
Now, there are several possible elections of the set $\Omega$.
Here we will consider only the set $Sep\,$ of separable states,
and the set $PPT$ of states with positive partial transposition.
Each of these choices leads to a different constraint. In
particular, Eq. (\ref{cond1}) becomes two conditions, \bea
E^{reg}_{Sep}(\rho_{AB}) = x,\\
E^{reg}_{PPT}(\rho_{AB}) = x. \eea We will next consider pure
states $\ket{\Psi_\delta}_{ABC}$ such that its reduced density
matrix for systems $AB$, $\delta$, is a PPT bound entangled state
\cite{bound}, and therefore $E^{reg}_{PPT}(\delta) = 0$. First we
will prove that $E^{reg}_{Sep}(\delta) > 0$, which leads to the
contradiction $0=x >0$, indicating that $\ket{\Psi_\delta}_{ABC}$
can not be reversibly generated with states of $G_3$ \cite{GPV}.
Notice that when applied to the PPT state $\delta$, Eq.
(\ref{cond2}) for $\Omega=PPT$ implies that $x_1 = x_2$
\cite{explain3}. We will also prove that \be \lim_{N \rightarrow
\infty} \frac{E_{Sep}(\delta^{\otimes N}\otimes[EPR]^{\otimes x_1
N})}{N} > x_1, \label{bound} \ee that by substitution in Eq.
(\ref{cond2}) for $\Omega=Sep\,$ implies that $x_2 > x_1$.
Therefore, we must have $x_1=x_2>x_1$, which is again a
contradiction, this time meaning that the states of $G_3$ can not
reversibly generate the state $\ket{\Psi_\delta}$ {\em and} states
of $G_3$.

We construct the three--partite states $\ket{\Psi_\delta}_{ABC}\in
\C^{d_A}\otimes\C^{d_B}\otimes \C^{d_C}$ as purifications of any
PPT bound--entangled state $\delta$ in $\C^{d_A}\otimes\C^{d_B}$
with no products vectors in its range, the so-called edge bound
entangled states \cite{edge}. Examples of these states can be
found in Refs. \cite{edge,bes}. In order to proceed, we need the
following result.

\vspace{1mm}

{\bf Theorem 1 \cite{DiVincenzo}:} Consider a projector $P$ onto a
subspace $V$ of $\C^{d_A}\otimes\C^{d_B}$ that does not contain
any product vector. A positive constant $\alpha$ exists such that
for all $N \geq 1$, \be \max_{\ket{a_N\otimes b_N}} \bra{a_N
\otimes b_N} P^{\otimes N} \ket{a_N \otimes b_N} \leq \alpha^N,
\ee where $\ket{a_N \otimes b_N} \in
\C^{d_A^{\,N}}\otimes\C^{d_B^{\,N}}$ denotes a product state.
\vspace{1mm}

\vspace{1mm}

{\em Proof:} $P$ fulfills the following properties: ($i$) Since
there are no product vectors in $V$, a positive number
$\alpha_1<1$ exists such that $\bra{a_1 \otimes b_1}\,P\,
\ket{\,a_1 \otimes b_1} \leq \alpha_1$ for all product vectors
\cite{Terhal}. ($ii$) A positive number $c > 0$ exists such that
$I+cP$ is separable \cite{ZHSL}. Then, the proof proceeds as for
the projector $P_b$ of Ref. \cite{VC} with
$\alpha\equiv(1+\alpha_1\,c)/(1+c)$. $\Box$

The following theorem provides us with a bound for the relative
entropy of entanglement with respect to the set $Sep\,$ and
together with theorem 1 is the key to the main result.

\vspace{1mm}

{\bf Theorem 2:} Let $P$ be the projector onto the support of a
mixed state $\rho_{AB}$ of a bipartite system
$\C^{d_A}\otimes\C^{d_B}$, let $\ket{a\otimes b} \in
\C^{d_A}\otimes\C^{d_B}$ denote a product vector and let $\beta$
be \be \beta \equiv \max_{\ket{a\otimes b}} \bra{a \otimes b} P
\ket{a \otimes b}. \ee The relative entropy of entanglement with
respect to separable states is bounded below by \be
E_{Sep}(\rho_{AB}) \geq - \log_2 \beta. \ee

\vspace{1mm}

{\em Proof:} Let $\sigma_{AB} \in Sep\,$ be the separable state
such that $E_{Sep}(\rho_{AB}) = S(\rho_{AB}||\sigma_{AB})$. The
quantum relative entropy can only decrease under a
trace--preserving completely positive map $\E$ \cite{book}. In
particular, let us consider \be \E(\tau) \equiv P\tau P +
(I-P)\tau (I-P). \ee We find \bea
S(\rho_{AB}||\sigma_{AB}) \geq S(\E(\rho_{AB})||\E(\sigma_{AB}))= \nonumber\\
\mbox{tr} (\rho_{AB}\log_2\rho_{AB}-
\rho_{AB}\log_2P\sigma_{AB}P), \eea where in the last step we have
used that $\rho_{AB}$ is invariant under $\E$ and that we can
ignore the contribution $(I-P)\sigma_{AB}(I-P)$ because its
support $I-P$ is orthogonal to $P$. Indeed, notice that for
positive operators $N, M_1$ and $M_2$, $\log (M_1\oplus M_2) =
\log M_1 \oplus \log M_2$, and therefore $\tr [(N\oplus 0) \log
(M_1\oplus M_2)] = \tr (N\log M_1)$. Define \bea
t\equiv \tr (P \sigma_{AB}),\\
\sigma'_{AB} \equiv \frac{1}{t} P \sigma_{AB} P. \eea Then,
because $\sigma_{AB}=\sum_i p_i \proj{a_i\otimes b_i}$ is a
separable state, we have that $t \leq \beta$. We finally obtain,
\bea S(\rho_{AB}||\sigma_{AB}) \geq \tr (\rho_{AB}\log_2
\frac{\rho_{AB}}{t\sigma'_{AB}}) = \nonumber\\
 - \log_2 t + S(\rho_{AB}||\sigma'_{AB}) \geq -\log_2 t \geq -\log_2\beta,
\eea where we have used that for positive operators $N,M$ and a
positive constant $k$, $\tr (N\log kM) = \tr (N\log M) + ( \tr N)
\log k$, and the positivity of the quantum relative entropy
\cite{book}. $\Box$

We only need to concatenate theorems 1 and 2 to find that for any
edge state $\delta$ \be E_{Sep}(\delta^{\otimes N}) \geq -\log_2
\alpha^N, \ee and therefore \be E^{reg}_{Sep}(\delta) \geq
-\log_2\alpha >0, \ee which disprove the initial conjecture for
$G_3$.

Notice that we can use this result and the inequalities
(\ref{inequalities}) to extend the irreversibility proved in
\cite{VC} to all the edge states. Indeed, we have $0 =
E^{reg}_{PPT}(\delta) < E^{reg}_{Sep}(\delta)$, and both
quantities are between the entanglement cost $E_c$ and the
distillable entanglement $E_d$.

Let us move now to prove Eq. (\ref{bound}). We need the following
two lemmas.

\vspace{1mm}

{\bf Lemma 1:} Let $P$ be a projector onto a subspace $V$ of
$\C^{d_A}\otimes \C^{d_B}$, and let $\ket{a\otimes
b}\in\C^{d_A}\otimes \C^{d_B}$ be a product state. Then \be
\max_{\ket{a\otimes b}} \bra{a\otimes b} P \ket{a \otimes b} =
\max_{\ket{\psi}\in V} \lambda_1(\psi), \label{2max} \ee where
$\lambda_1(\psi)$ denotes the largest coefficient $\lambda_i$ in
the Schmidt decomposition of $\ket{\psi}$, $\ket{\psi}=\sum_i
\sqrt{\lambda_i} \ket{u_i\otimes v_i}$, $\lambda_1 \geq
\lambda_{i+1}$.

\vspace{1mm}

{\em Proof:} For any product vector $\ket{a\otimes b}$, let us
define the normalized vector $\ket{\gamma}\in V$ as
$P\ket{a\otimes b}/||P\ket{a\otimes b}||$. Then \be
 \bra{a\otimes b} P \ket{a \otimes b} =
  |\braket{a\otimes b}{\gamma}|^2 \leq \lambda_1(\gamma),
\ee where in the last step we have used lemma 1 of \cite{VJN}. Let
$\ket{\psi'}$ be the vector for which the maximum in the rhs of
Eq. (\ref{2max}) is attained, and let $\sum_i \sqrt{\lambda'_i}
\ket{u'_i\otimes v'_i}$, $\lambda_i'\geq \lambda'_{i+1}$, be its
Schmidt decomposition. Then \be \max_{\ket{\psi}\in V}
\lambda_1(\psi)= \lambda'_1 = \bra{u'_1\otimes v'_1} P
\ket{u'_1\otimes v'_1}, \ee which finishes the proof. $\Box$

\vspace{1mm}

{\bf Lemma 2:} Let $P$ be a projector onto a subspace $V$ of
$\C^{d_A}\otimes \C^{d_B}$ and let $P_{\Phi}$ be a projector onto
a bipartite pure state $\ket{\Phi}\in \C^{d'}\otimes \C^{d'}$ with
Schmidt decomposition $\sum_{i=1}^{d'} \sqrt{\lambda_i} \
\ket{u_i}\otimes \ket{v_i} $, $\lambda_i\geq\lambda_{i+1}$.
Finally, let $\alpha_p$ be \be \alpha_p \equiv \max_{\ket{a\otimes
b}} \ \bra{a\otimes b} \ P \ \ket{a\otimes b}, \ee where
$\ket{a\otimes b} \in \C^{d_A}\otimes \C^{d_B}$ denotes a product
state. Then, \be \max_{\ket{c\otimes d}} \ \ \bra{c\otimes d} \ \
P\otimes P_{\Phi} \ \ \ket{c\otimes d} \, =  \, \alpha_p\lambda_1,
\label{lemma2} \ee where the maximization is made over product
vectors $\ket{c\otimes d} \in \C^{d_A+d'}\otimes\C^{d_B+d'}$.

\vspace{1mm}

{\em Proof:} Notice that $P\otimes P_{\Phi}$ projects onto a
subspace spanned by vectors of the form
$\ket{\psi}\otimes\ket{\Phi}$, $\ket{\psi} \in V$, and that the
largest coefficient $\lambda_1$ in a Schmidt decomposition
fulfills $\lambda_1(\psi\otimes\Phi) =
\lambda_1(\psi)\lambda_1(\Phi)$. Then Eq. (\ref{lemma2}) follows
from lemma 1. $\Box$

\vspace{1mm}

We would like to bound below the relative entropy of entanglement
$E_{Sep}$ of \be \delta^{\otimes N} \otimes [EPR]^{\otimes M}. \ee
The projector onto its support is given by $P_\delta^{\otimes N}
\otimes [EPR]^{\otimes M}$, where $P_\delta$ is the projector onto
the support of $\delta$, and we can use lemma 2 and theorem 1 to
obtain \be \max_{\ket{a\otimes b}} \bra{a\otimes b}
P_\delta^{\otimes N} \otimes [EPR]^{\otimes M} \ket{a\otimes b}
\leq \frac{\alpha^N}{2^M}, \ee where $(1/2)^M$ corresponds to
$\lambda_1(EPR^{\otimes M})$. Then we can apply theorem 2 to
obtain \be E_{Sep} (\delta^{\otimes N} \otimes [EPR]^{\otimes M})
\geq -N\log_2 \alpha + M, \ee which implies Eq. (\ref{bound}).
This finishes the proof of the fact that it is not possible to
reversibly transform states of $G_3$ into any purification of a
PPT edge state and states of $G_3$.

%In this work we have showed by means of counter--examples that GHZ and EPR states alone cannot be used to reversibly generate all three--partite pure states.This result leaves several questions open.
It would be interesting to understand the mechanisms that lead to
this irreversibility. Recall that in the asymptotic limit some
non-trivial three--partite states can be reversibly generated from
EPR and GHZ states \cite{VDC}. We ignore which conditions
determine whether a three--partite pure--state transformation can
be performed in a reversible way. The following two facts
suggest, however, that there may be a connection between this
question and the irreversibility that takes place during the
preparation--distillation cycle of bipartite mixed states:

($i$) All known three-partite reversible transformations
\cite{BPRST,VDC} involve pure states whose bipartite reduced mixed
states can be distilled and prepared in a reversible way
\cite{explain4}.

($ii$) The proof that $G_3$ is not a MREGS relies on the
irreversibility that occurs in bipartite mixed--state
manipulation. Indeed, suppose that $E_c$ and $E_d$ would not
disagree for edge states. Then, because of Eq.
(\ref{inequalities}), $E^{reg}_{PPT}$ and $E^{reg}_{Sep}$ would
also have been equal, and this would jeopardize our argument.

Finally, a major open question is whether a finite MREGS exists
for three-partite states and, if so, which kind of states must
include. These are difficult issues that certainly deserve further
investigation. We cautiously conclude the present work by noting
that the states of an eventual MREGS must have bipartite reduced
density matrices able to reproduce the discrepancies between
relative entropies displayed by the states $\delta$, and must
therefore carry themselves the signature of bipartite mixed--state
irreversibility.

%\section{Acknowledgements}

A. A. thanks J. Preskill and the IQI, Caltech, for hospitality. We
thank D. P. DiVincenzo, W. D\"ur, E. Jan\'e, N. Linden, Ll.
Masanes and S. Popescu for discussion. This work was supported by
the European project EQUIP (IST-1999-11053), by the ESF, by the
Swiss FNRS and OFES, and by the NSF (of USA), Grant. No.
EIA-0086038.

%------------------ uuuuuuu!
Note added: after completion of this work, Y. Shi pointed out the
relation between the results proved here and his recent work
\cite{Shi}. We have not been able to follow the line of
argumentation in such a work.

\bigskip

\small{$^*$Antonio.Acin@physics.unige.ch}

\small{$^\dagger$vidal@cs.caltech.edu}

\small{$^\ddagger$Ignacio.Cirac@mpq.mpg.de}


\begin{references}
\bibitem{explain2}
The relative entropy of entanglement  fulfills
$E_{\Omega}(\rho_{AB}\otimes \rho_s)=E_{\Omega}(\rho_{AB})$ for
any separable state $\rho_s$ as a result of its monotonicity under
LOCC \cite{relent}.
\bibitem{explain3}
For $\rho_{ppt}$ a PPT state we have $E_{PPT}(\rho_{AB}\otimes
\rho_{ppt})=E_{PPT}(\rho_{AB})$, since ($i$) by means of LOCC we
can get rid of $\rho_{ppt}$ and LOCC can only decrease $E_{PPT}$,
so that $E_{PPT}(\rho_{AB}\otimes \rho_{ppt}) \geq
E_{PPT}(\rho_{AB})$, and ($ii$) for any PPT state $\pi_{ppt}$ we
have $S(\rho_{AB}\otimes\rho_{ppt}||\pi_{ppt}\otimes\rho_{ppt}) =
S(\rho_{AB}||\pi_{ppt})$, which guarantees that
$E_{PPT}(\rho_{AB}\otimes \rho_{ppt})$ is not going to be larger
than $E_{PPT}(\rho_{AB})$.
\end{references}
\end{document}